\documentclass{article}
\usepackage[english]{babel}
\usepackage{amsfonts}
\usepackage[letterpaper,top=2cm,bottom=2cm,left=3cm,right=3cm,marginparwidth=1.75cm]{geometry}
\usepackage{amsmath}
\usepackage{graphicx}
\usepackage{geometry}
\usepackage{physics}
\usepackage[colorlinks=true, allcolors=blue]{hyperref}
\usepackage{authblk}

\title{On the Derivation of Equations of Motion from Symmetries in Quantum-Mechanical Systems via Heisenberg's Uncertainty Principle}   
\author[1]{Enrique Casanova}
\author[1]{José Rojas}
\author[1,2]{Melvin Arias}

\affil[1]{\textit{\small Instituto de Física, Universidad Autónoma de Santo Domingo, Av. Alma Mater, Santo Domingo 10105, Dominican Republic}}
\affil[2]{\textit{\small Laboratorio de Nanotecnología, Área de Ciencias Básicas y Ambientales, Instituto Tecnológico de Santo Domingo, Av. Los Próceres, Santo Domingo 10602, Dominican Republic}}

\begin{document}
\maketitle

 \begin{abstract}
We propose the construction of equations of motion based on symmetries in quantum-mechanical systems, using Heisenberg’s uncertainty principle as a minimal foundation. From canonical operators, two spaces of conjugate operators are constructed, along with a third space derived from the former, which includes the “Symmetry-Dilation” operator. When this operator commutes with the main equation of motion, it defines the set of observables compatible with a complete basis of operators (symmetry generators), organized into a Lie algebra dependent on Heisenberg’s uncertainty principle within Minkowski spacetime. Furthermore, by requiring the dilation operator to commute with the central operator, the wavefunction is constrained, thereby constructing known structures. Specific cases are derived—relativistic, non-relativistic, and a lesser-studied case: “ultra-relativistic (Carroll-Schrödinger).” Our work may open new avenues for understanding and classifying symmetries in quantum mechanics, as well as offer an alternative method for deriving equations of motion and applying them to complex scenarios involving exotic particles.

\end{abstract}

\section{Introduction} 
Since the 20th century, the equations of motion in quantum mechanics have been derived either through the principle of least action or from algebraic postulates, as in the formulations of Schrödinger, Heisenberg, and Dirac. Although these formulations are dynamically equivalent, they present different constructions and motivations. A significant advance in algebraic unification and formalization was made by John von Neumann in 1955, through his work in Mathematical Foundations of Quantum Mechanics, where he developed abstract algebras based on Hilbert spaces and bounded operators, laying the foundation for a consistent mathematical structure. On the other hand, the development of quantum field theory (QFT) re-established the Lagrangian formulation as a bridge to derive equations of motion from the principle of least action, resolving the problems in relativistic quantum mechanics. However,
despite the extensive use of Lagrangian and Hamiltonian approaches, there is no systematic framework for deriving equations of motion directly from algebraic symmetry constraints, the choice of a Lagrangian density depends on various properties such as energy scale and experimentally adjusted parameters, as well as theoretical principles like relativity and gauge invariance. These latter are properties closely tied to the system’s symmetries, which can be obtained through different methods (as in Noether’s theorem), and their importance lies in the transformations that leave certain physical quantities (observables) of the system invariant \cite{Wigner1931}. Simultaneously, with the fulfilment of Heisenberg’s uncertainty principle, which limits the measurement of certain conjugate pairs of physical properties, and which was formalized by Kennard \cite{Kennard} and Weyl \cite{Weyl1928}, and later linked by Ozawa to a general conjecture involving errors and disturbances \cite{Ozawa2002}\cite{Ozawa2003}\cite{Woods2013}. It is used as a foundation in quantum metrology, a branch of physics that uses quantum principles to achieve high-accuracy and precision measurements, creating and manipulating entangled states \cite{GiovannettiLloydMaccone2011}\cite{ZhuangLee2024}\cite{HuangLee2024}.

In this work, we present a symmetry-based method that derives equations of motion for free quantum particles without resorting to variational principles. First, imposing the algebraic structure of symmetry generators as a foundation, and assuming the Heisenberg uncertainty principle as a minimal and intrinsically valid postulate. We are inspired by recent works on the construction of Casimir invariants \cite{AlshammariIsaacMarquette2018}\cite{EswaraRao2021}\cite{Cantuba2021}. Nevertheless, we highlight the seminal work of Lubański, \textit{On the theory of elementary particles of arbitrary spin. I–II.} \cite{Lubanski1942}, as one of the earliest to employ symmetries (the translation generator and the rotation/boost generator) to construct an alternative Casimir invariant consistent with the Poincaré group. 

Beginning with the theoretical framework in Section 2, which is based on the definition of an operator $\hat{Q}$, \textit{symmetry-dilation}, in an operator space, expanded in the symmetry basis, which satisfies the Lie algebra of symmetry generators via the HUP, of the central operator $\hat{A}$ (Casimir) belonging to the centralizer $\mathcal{Z}(\hat{Q})$. 
The explicit construction of the equations of motion arises from restricting the central operator after commuting with the operator we define as “dilation”, proposing the system as scale-invariant, thus restricting the wave functions. In Section 3, we obtain structures in our operator space of relativistic, non-relativistic, and a special “ultra-relativistic” case (Carroll–Schrödinger typology \cite{Najafizadeh2025}, which is normally obtained from the Klein–Gordon equation by substitution and then approximating 
$c \xrightarrow{} 0$) with rotational symmetry breaking in the scalar case, but with “ultra-boost” symmetry in Minkowski spacetime, being quadratic in the temporal translation operator and linear in the spatial translation operators, potentially connecting to cosmological phenomena by coupling to Carroll symmetries \cite{Bergshoeff2014}\cite{LevyLeblond1965}\cite{deBoer2022}, or even as a candidate for dark matter \cite{Najafizadeh2024}. Unlike Galilean or Lorentzian boosts, our ultra-boost operator inverts conventional notions of symmetric transformation, offering a perspective on (anti-)unitary evolution of quantum systems in non-standard regimes. To recover the system’s isotropy, in Section 3.2.3, we construct a framework similar to the relativistic Dirac case with gamma matrices (end of Section 3.1), but using Pauli matrices. Their coupling to the usual rotation operator restores the system’s isotropy, but imposes a mandatory matrix representation, thus mapping to a distinct representation with the same structure as the Carroll–Schrödinger equation and creating a notion of spin for this case in the non-standard wave dispersion regime  $\omega \propto \sqrt{|\mathbf{k}|}$. Finally, by using the same construction of the Pauli equation from the non-relativistic case, via a Galilean boost transformation (Section 3.2.1), but for the ultra-relativistic matrix equation with the ultra-boost, we derive the Carroll–Schrödinger–Pauli equation analogous to the Pauli equation with potentials $\phi_i$ coupled to the momentum operators $\hat{p}_i$ in a non-linear way. 

Our approach unifies relativistic, non-relativistic, and ultra-relativistic (Carroll–Schrödinger type) regimes within a single operator framework, and naturally classifies the basis of the set of compatible observables via the centralizer of our symmetry–dilation operator.

\section{Theoretical Framework}
First, the extended commutation relations in phase space (operators acting on $L^2(\mathbb{R}^4)$) in $(\hat{x}_\mu, \hat{p}_v) \in \Omega(\hat{x}_\mu,\hat{p}_v)$ are: 
\begin{equation}
\begin{aligned}
\comm{\hat{x}_\mu}{\hat{x}_\nu} &= i\hbar \theta_{\mu\nu} \\
\comm{\hat{p}_\mu}{\hat{p}_\nu} &= i\hbar \Sigma_{\mu\nu} \\
\comm{\hat{x}_\mu}{\hat{p}_\nu} &= i\hbar g_{\mu\nu}
\end{aligned}
\end{equation}

where $\theta_{\mu v}$, $\Sigma_{\mu v}$ and $g_{\mu v}$ are functions in  $f(\Omega)$. If these functions are constant, quantum non-locality is also ensured (see \cite{OppenheimWehner2010}). 

Defining next the following operators: $\hat{N}_\alpha$ and $\hat{M}_\beta$ as functions of $\hat{x}_\mu$ and $\hat{p}_\mu$ in $f(\Omega)$ as: 
\newline 

Operator \textit{"Position"} (\textit{Dual Translation}): \begin{equation} \hat{N}_\alpha  \equiv f_\alpha(\hat{x}_\mu) \end{equation}

Operator \textit{"Dual Position"} (\textit{Translation}): \begin{equation} \hat{M}_\beta \equiv f_\beta(\hat{p}_\nu) \end{equation}

with $\hat{x}_0 = c\hat{t}$ and $\hat{p}_{0}=\frac{\hat{H}}{c}$. 
the first operator forms a basis for a operator space $Z(\hat{N}_\alpha)$ of \textit{"Dual Translation"}, (i.e $\hat{N}_\alpha \in Z(\hat{N}_{\alpha}) \subset f(\Omega) $), the second operator also forms a basis but, in a operator space $Z(\hat{M}_\beta)$ of \textit{"Translation"}, (i.e $\hat{M}_\beta \in Z(\hat{M}_{\beta}) \subset f(\Omega)$). These two operator spaces each form an \textit{operator algebra} with unit element $\hat{I}$. They are often equipped with their own topologies, for example: norm topology, weak topology, weak* topology, etc.
A special class of C*-algebras are the Von Neumann algebras, which (in the complex case) are weakly closed *-subalgebras (closed in the weak topology) of the algebra of bounded linear operators $B(H)$ \cite{Sakai1971}. 
We will assume this case throughout the article, with wave-functions $\ket{\Phi} (x^\mu) \in  L^2(\mathbf{R}^4)$ in a finite complex Hilbert space with four degrees of freedom (three spatial and one temporal). The choice of a Von Neumann algebra (type I) ensures consistency with relativistic quantum mechanics, allowing the construction of symmetry operators with proper physical interpretation. Finally, although the explicit construction of the wave-function's expectation value is not addressed in the article, it can be derived from the state density outlined in Von Neumann theory.
Now, we construct a third operator space from the previous ones:
\begin{equation}
    Z_{\alpha \beta} \equiv Z(\hat{N}_{\alpha}) \times Z(\hat{M}_{\beta})  
\end{equation}
This gives us the possibility to define a function $F: Z_{\alpha \beta} \xrightarrow{} f(\Omega) $, with the closed operation: 
\begin{equation}
     F( \hat{N}_\alpha, \hat{M}_\beta) \equiv\hat{N}_\alpha \hat{M}_\beta = \frac{1}{2} \comm{\hat{N}_\alpha }{\hat{M}_\beta } + \frac{1}{2}  \{ \hat{N}_\alpha , \hat{M}_\beta \} 
 \end{equation}
The space $F(Z_{\alpha \beta})$ forms an associative algebra of bounded operators generated by the operation, where both $Z(\hat{N}_{\alpha})$ and $Z(\hat{M}_{\beta})$ belong to this algebra. From this operation, we observe that it contains an antisymmetric part (commutator), which is Lie-admissible, and a symmetric part (anti-commutator), which is Jordan-admissible, making it a non-Abelian operation. If we take the commutator to be equal to $i\hbar C_{\alpha \beta}$, where $ C_{\alpha \beta}$ is a parameter dependent on definitions (2) and (3), we obtain the canonical quantization condition for conjugate operators in the space $Z_{\alpha \beta}$, which directly connects to the usual Bose–Einstein statistics. Similarly, we can map to Fermi–Dirac statistics by using the anti-commutator in equation (5) instead of the commutator.

\subsection{Symmetry-Dilation Factorization}

The following construction arises from assuming the minimal structure of a quantum system: the Heisenberg–Weyl uncertainty relation, which dictates an intrinsic connection between the “Symmetry-Dilation” operator" \( \hat{Q} \in F(Z_{\alpha \beta})\) and the Casimir operator \( \hat{A} \in  F(Z_{\alpha \beta}) \). Consequently, the dynamical equation governing the wavefunction of the quantum system emerges from imposing: \begin{equation}
    \comm{\hat{A}}{\hat{Q}} = 0
\end{equation}

thus $\hat{A} \in \mathcal{Z}( \hat{Q})$ meaning it belongs to the centralizer $\hat{Q}$. This commutation condition guarantees the theorem of compatible observables (CSCO) for $\hat{A}$ —\textit{first discussed by Dirac in P. A. M. (1930): The Principles of Quantum Mechanics}-. In this way, we ensure that the operators are simultaneously measurable within the system and that their solutions share the same Hilbert space $L^2$. Moreover, by understanding how the base operators are expanded and organized within $\hat{Q}$, we can determine how many symmetries the quantum system admits.

\subsection{Hypotheses}

\begin{itemize}
    \item Canonical algebra::
    \[
    \comm{\hat{x}_\mu}{\hat{p}_v} = i\hbar \eta_{\mu v}
    \]
    compactified in four-vector form using the Minkowski metric, $\eta_{\mu v} = diag(-1,+1,+1,+1)$ .
    \item Minimal linear operators, from (2) and (3):
    \begin{itemize}
        \item Linear Dual Translation:
        \[
        \hat{N}_\alpha = \alpha_\mu \hat{x}^\mu + \alpha' \hat{I},
        \]
        \item Linear Translation:
        \[
        \hat{M}_\beta = \beta_\nu \hat{p}^\nu + \beta' \hat{I},
        \]
       (using Einstein summation convention over \( \mu, \nu = 0, 1, 2, 3 \)).
    \end{itemize} 
    
   \item Taking $\hat{Q}(\hat{x}_\mu,\hat{p}_v) \equiv \hat{N}_\alpha \hat{M}_\beta$, the self-adjoint symmetry operator is factorized as a complete combination of symmetry generators and the “dilation” operator:
    \[
    \hat{Q} = c_l \hat{I} + c_p^\mu \hat{p}_\mu + c_x^\mu \hat{x}_\mu + c_m^{\mu\nu} \hat{M}_{\mu\nu} + \hat{\mathcal{D}}  = \hat{\mathcal{S}} + \hat{\mathcal{D}}
    \] 
    where the coefficients \( c \) are polynomial functions of \( \alpha_\mu, \beta_\nu, \alpha', \beta' \). The symbols \( \hat{M}_{\mu \nu},\hat{\mathcal{S}} \)  y \(\hat{\mathcal{D}} \) represent, respectively: rotations/boosts, symmetries, and dilation.  
    \item Symmetry Condition (Casimir): The existence of a Hermitian operator \( \hat{A} \) is required such that it commutes with $\hat{Q}$ for some $\ket{\Phi_c} \in L^2(\mathbf{R}^4)$. 
    
\end{itemize}
\subsection{Affirmations}

\begin{enumerate}
    \item[A1]: \textit{Determination of \( \hat{A} \)}: Under the previous hypotheses, the condition \( [\hat{A}, \hat{\mathcal{S}}]\ket{\Phi} = 0 \), uniquely determines \( \hat{A} \) as a function of the symmetry generators, in general:
    \[
    \hat{A} \equiv G(\hat{\mathcal{S}}) = \sum^\infty_{n=0} q_n {\hat{\mathcal{S}}}^n
    \]
    where the coefficients \( q_n \) are real numbers (or real matrices). Furthermore, in some applications, the Casimir may be a function of a specific symmetry of $\hat{\mathcal{S}}$.
    
    \item[A2]: \textit{Universal Dynamical Equation}: For some state vector $\ket{\Phi_c}$, the following condition must hold:
    \[
    \comm{\hat{A}}{\hat{\mathcal{D}}} \ket{\Phi_c} = 0 \sim \hat{A} \ket{\Phi_c} = 0
    \]
    where \( \hat{\mathcal{D}} \) is the dilation operator, explicitly defining it as:
    \[\hat{\mathcal{D}} =  \alpha^\mu \beta^\mu   \hat{x}_\mu \hat{p}_\mu \]
        
\end{enumerate}
Remarking that, in the development of quantum field theory, the analysis of scale symmetries has been systematically addressed through the dilation  generator, which in four dimensions was defined by Peskin and Schroeder \cite{Peskin1995}, used to determine the canonical dimensions of fields and study scale anomalies. Later, in the context of two-dimensional conformal theories, Di Francesco et al. \cite{DiFrancesco1997}, with the Virasoro algebra, emphasized its central role in the classification of primary fields and in the conformal structure.

\section{Derivation Cases: Relativistic, Non-Relativistic, and Ultra-Relativistic} 

All subsequent subsections are obtained as particular application, assuming that $\hat{A} = G(c^\mu_p \hat{p}_\mu)$ is only a function of $ \hat{p}_\mu$. It suffices to group the elements within $\hat{N}_\alpha \hat{M}_\beta$ so that they generate the desired algebra, thereby expressing different cases of $\hat{Q}$. Then \( \hat{A} \) is computed using the commutation rules, ensuring the fulfilment of equation (6), and the condition \( \comm{\hat{A}}{\hat{\mathcal{D}}} \ket{\Phi_c} = 0 \), is imposed, thereby satisfying the assertions of Section 2.3. In all these cases, the notion of spin as an internal symmetry arises from the definition of the elements $\alpha_n$ and $\beta_n$, which themselves satisfy a specific algebra. 

\subsection{Relativistic Equations: Dirac and Klein–Gordon}

In the case of grouping relativistic symmetries imposed on the operator \(\hat Q\), its commutation condition with $\hat{A}$ is of adjustable degree $k$ (i.e $\hat{p}^k_\mu$), By choosing it to be quadratic, the second-order Klein–Gordon equation for integer spins emerges immediately. If taken to be linear, the first-order Dirac equation for half-integer spins simultaneously arises. Thus, both formulations are derived from a common symmetric core and are mutually implied within the same algebraic framework.
    
Grouping of Rotations/Boosts 
\begin{equation}
        \alpha_\mu \beta_v = -\alpha_v \beta_\mu
\end{equation} 

The relativistic operator $\hat{\mathcal{S}}$ is now written as a linear combination of the symmetry generators: 

\begin{equation}
    \hat{\mathcal{S}} = \alpha' \beta' +\alpha^\mu \beta^v \hat{M}_{\mu v} + \alpha^`\beta^\mu \hat{p}_\mu + \alpha^\mu\beta' \hat{x}_\mu 
\end{equation}
Where $\hat{M}_{\mu v} = \hat{x}_\mu \hat{p}_v - \hat{x}_v \hat{p}_\mu $, are the Lorentz rotations and boosts in the operator representation. 
Note that the first ten operators satisfy the commutation relations characteristic of the Lie algebra of the Poincaré group, in the Klein–Gordon case, as generators of continuous symmetries. For the Dirac case, it is necessary to impose the condition $\alpha' \beta' = k_0 \alpha^\mu \beta^v \sigma_{\mu v} $, where $\sigma_{\mu v} = -\sigma_{v \mu}$, is antisymmetric, so as not to break the Lie algebra structure and instead couple to $\hat{M}_{\mu v} \xrightarrow{} \hat{M}_{\mu v} + k_0\sigma_{\mu v}$. Finally, we observe an additional symmetry from the last element of (8); this symmetry acts as a constraint that will acquire physical meaning later. Now, taking the central operator, 

\begin{equation}
    \hat{A} = \sum^\infty_{n=0} a_n (\alpha' \beta^\rho \hat{p}_\rho)^n 
\end{equation}

which must commute with each rotation and boost of the system $\comm{\hat{A}}{\alpha' \beta' +\alpha_\mu \beta_v \hat{M}_{\mu v}} = 0$, and $a_{n>1} \in \mathbb{R}$, and $a_0$ could be a constant matrix. From this, we obtain the first constraint on the central operator: 
\begin{equation}
    \alpha_\mu \beta_v \comm{\hat{A}}{\hat{M}_{\mu v} + k_0\sigma_{\mu v}} = -\comm{\hat{A}} {\alpha_\mu\beta_v} (\hat{M}_{\mu v} + k_0\sigma_{\mu v})
\end{equation}
The commutator on the left-hand side must vanish for the operator to satisfy the full Poincaré algebra with spin. Therefore, the right-hand commutator, which constrains the internal symmetries, must also be zero. Expressing the first commutation explicitly:

\begin{equation}
    \comm{(\alpha' \beta^\rho \hat{p}_\rho)^n}{\hat{M}_{\mu v}} = -k_0 \comm{(\alpha' \beta^\rho \hat{p}_\rho)^n}{\sigma_{\mu v}}
\end{equation}

We highlight that when $n=2k$ (even), the double commutation holds $\comm{(\alpha' \beta^\rho \hat{p}_\rho)^{2k}}{\hat{M}_{\mu v}} = 0 \xrightarrow[]{} \comm{(\alpha' \beta^\rho \hat{p}_\rho)^{2k}}{\sigma_{\mu v}} = 0 $. For example, by choosing $k = 1$, we obtain the first Klein–Gordon relation: $(\alpha'\beta^c) (\alpha' \beta_d) \comm{\hat{p}_c \hat{p}^d}{\hat{M}_{\mu v}}=0$, which, since $\comm{\hat{p}_\sigma \hat{p}^\sigma}{\hat{M}_{\mu v}}$ commutes, must be taken accordingly,
\begin{equation}
    \{ \alpha'\beta_\mu , {\alpha' \beta_v} \} = 0 \xrightarrow[]{} \comm{(\alpha' \beta^\rho)^{2}}{\sigma_{\mu v}} =0
\end{equation}

when $(\mu \neq v)$,the summation in (9) is significantly simplified by removing the dependence on the cross terms $\hat{p}_\mu \hat{p}_v ...$, bringing us closer to the usual structure of a relativistic operator in Minkowski space. Now, equation (11) becomes:
\begin{equation}
    (\alpha' \beta^\rho)^n \comm{\hat{p}^n_\rho}{\hat{M}_{\mu v}} = -k_0 \comm{(\alpha' \beta^\rho)^n}{\sigma_{\mu v}}\hat{p}_\rho  
\end{equation}
As previously discussed, depending on whether $n$ is even or odd, we obtain two distinct commutation constraints for each case. Continuing from (10), we finally arrive at the constraint on the elements, $\comm{(\alpha' \beta_\rho)}{\alpha_{\mu} \beta_v} = 0$ (which, as required for $n=1$ in the Dirac case, holds for any value of $n$).
Finally, the commutation of $\hat{A}$ with the summation $\alpha^\mu \beta' \hat{x}_\mu$ yields: 
\begin{equation}
    \comm{\alpha^\mu \beta'}{\sum a_n(\alpha' \beta^\rho)^n} \hat{x}_\mu \hat{p}_\rho + i\hbar \sum a_n(\alpha' \beta^\rho)^n (\alpha^\mu \beta') \eta_{\mu \rho} = 0
\end{equation} From all this, the candidate for $\sigma_{\mu v}$ — due to its antisymmetry and the relations in (12) — is
\begin{equation}
    \sigma_{\mu v} = \frac{i}{2}\comm{\alpha' \beta_\mu }{\alpha' \beta_v}
\end{equation}
with $k_0 = -\hbar$. Equation (15) thus becomes the standard representation of the spin group, provided the terms  $\alpha' \beta_\mu$ are taken as the Dirac gamma matrices $\gamma_\mu$. Thus, from equation (9) we extract the two operators of interest:
\begin{enumerate}
    \item[] Minimal odd case (Dirac-type equation): \begin{equation}
         \hat{A}_{D} = a_1\gamma^\rho \hat{p}_{\rho}
    \end{equation} 
    \item[] Minimal even case (Klein–Gordon-type equation): \begin{equation}
        \hat{A}_{KG} = - a_2\hat{p}^\rho \hat{p}_\rho
    \end{equation} 
\end{enumerate}

Note that by taking $\alpha^\mu \beta' = \gamma^\mu$ equation (14) leads to:
\begin{enumerate}
    \item[] $n = 1$:\begin{equation}
        \sigma^{\mu v} \hat{M}_{\mu v} + 4\hbar I_4 = 0
    \end{equation}
\end{enumerate}
with $I_4$ as the identity matrix in four dimensions. For the case $n=2$ it holds that $a_0 - a_2 \eta^{\rho \rho}=0$. As for the orders $n>3$, different internal relations will arise.

Equations (16) and (17) must satisfy Statement 2 from Section 2.3. We have:
\begin{equation}
     \comm{\hat{A}_{KG}}{\hat{\mathcal{D}}} \ket{\Phi_c} = 0 \longrightarrow  [\alpha_\mu \beta_\mu \hat{p}^\mu \hat{p}^\mu] (\mathbf{1}\ket{\Phi_c}) = 0 
\end{equation} 
\begin{equation}
    \comm{\hat{A}_{D}}{\hat{\mathcal{D}}} \ket{\Phi_c} = 0 \longrightarrow  [\alpha_\mu \beta_\mu \gamma_\mu \hat{p}^\mu] (\mathbf{1}\ket{\Phi_c}) = 0
\end{equation}
These are the expected equations of motion for our relativistic system. In the case where $\alpha_\mu \beta_\mu  = \mathbf{1}$ (unitary representation), we recover the usual Klein–Gordon and Weyl equations, respectively.

\subsection{Equations: Non-Relativistic and Ultra-Relativistic}

We now assume a different grouping of the internal elements of $\hat{Q}$, one that approximates the ten usual non-relativistic symmetries: rotations, translations, and Galilean boosts. By obtaining the explicit representation of these three Galilean symmetries, we also derive a representation of the three ultra-relativistic boosts. These correspond to quantum systems where either the speed of light is assumed to be very small, or the system in question moves faster than the speed of light. It is possible to separate the symmetry operator into two parts now: 
\begin{equation}
    \hat{S} = \hat{S}^{NR} + \hat{S}^{UR}
\end{equation}
where the grouping of rotations follows the usual form: $\alpha_i \beta_j = - \alpha_j \beta_i$. Now, starting with the substitution: $\beta' \xrightarrow[]{} \beta'/c^2$ and,

\begin{equation}
\begin{aligned}
\beta^0 \xrightarrow[]{} c\beta^0  \\
\alpha^0 \xrightarrow[]{} \alpha^0/c \\
\end{aligned}
\end{equation}
We obtain:
\begin{enumerate}
     
    \item[] Grouping of non-relativistic boosts: 
        \begin{equation}
            \alpha^{0} \beta^i = -(2a_{NR}) \alpha^i \beta^`
        \end{equation}
    \item[] Grouping of ultra-relativistic boosts:  
        \begin{equation}
            \alpha^{i} \beta^0 = \frac{(2a_{UR})}{c^2} \alpha^0 \beta^` 
        \end{equation}
    
\end{enumerate}
where $a_{NR}$ and $a_{UR}$ are real scalars (not dimensionless). The latter constant can be made dependent on the square of a velocity adjustment parameter (with the same units as velocity), Therefore, by setting $a_{UR} = v^2a^*_{UR}$we obtain from equation (24) the transformation on the right-hand side: $\frac{2v^2a^*_{UR}}{c^2}$. 

Thus $c > > v$, we approximate:
\begin{equation}
   \hat{\mathcal{S}} \xrightarrow{} \hat{\mathcal{S}}^{NR} = \alpha^i \beta^j\hat{M}_{i j} + \alpha^{i} \beta' \hat{B}^{NR}_i + \alpha^` \beta^i \hat{p}_i - \alpha'\beta_0 \hat{H}
\end{equation}

where $\hat{B}^{NR}_i =[(2a_{NR})\hat{t}\hat{p}_i+\hat{x}_i]$ are the Galilean boost operators (non-relativistic boosts). These generators exhibit similarities with the Galilean Conformal Symmetry (GCS), which finds applications in boundary field theories and in AdS/CFT correspondence \cite{Bagchi2009}. 
On the other hand, if $c < < v$, we approximate to:
\begin{equation}
   \hat{\mathcal{S}} \xrightarrow{} \hat{\mathcal{S}}^{UR} = (\alpha^i \beta^j\hat{M}_{i j} + \alpha' \beta'/c^2) - \alpha_{0} \beta' \sum_i  \hat{B}^{UR}_i + \alpha^` \beta^i \hat{p}_i - \alpha'\beta_0 \hat{H}
\end{equation}

Where $\hat{B}^{UR}_i = [\frac{v^2}{c^2} (2a^*_{UR})\hat{x}_i \hat{H} + t]$. These types of approximations involving a small speed of light (in comparison) have already been studied in: \cite{Bergshoeff2014} \cite{deBoer2022} \cite{LevyLeblond1965} and \cite{Najafizadeh2025}. Since this algebra differs significantly from the usual Lie algebras (Poincaré and Galilean), we now write the algebra of external symmetry generators: 

\begin{enumerate}
    \item[] Spatial Translations / Spatial Translations \begin{equation}
        \comm{\hat{p}_i}{ \hat{p}_j} = 0 
    \end{equation}
    \item[] Spatial Translations / Temporal Translation
        \begin{equation}
            \comm{\hat{p}_i}{\hat{H}} = 0
        \end{equation}
    \item[] Spatial Translations / Rotations
        \begin{equation}
            \comm{\hat{p}_k}{\hat{M}_{i j}} =  i\hbar( \delta_{j k} \hat{p}_i-\delta_{i k} \hat{p}_j) 
        \end{equation}
    
    \item[] Rotations / Rotations 
     \begin{equation}
        \comm{\hat{M}_{i j}}{\hat{M}_{k h}} =  i\hbar (\delta_{i k} \hat{M}_{j h} - \delta_{i h} \hat{M}_{j k} - \delta_{j k} \hat{M}_{i h} + \delta_{j h} \hat{M}_{i k})
     \end{equation}
    \item[] Rotations / Temporal Translation \begin{equation}
         \comm{\hat{M}_{i j}}{\hat{H}} = 0
    \end{equation}
    \item[] Rotations / Ultra-Relativistic Boosts
        \begin{equation}
            \comm{\hat{M}_{i j}}{\hat{B}^{UR}_k} = -2i\hbar a_{UR} (\frac{v^2}{c^2}) [\delta_{j k} \hat{x}_i - \delta_{i k}\hat{x}_j] \hat{H} 
        \end{equation}
    \item[] Ultra-Relativistic Boosts / Spatial Translations \begin{equation}
        \comm{\hat{B}^{UR}_i}{\hat{p}_j} = 2i\hbar a_{UR}(\frac{v^2}{c^2})\delta_{i j}\hat{H}
    \end{equation}
    \item[] Ultra-Relativistic Boosts / Temporal Translation \begin{equation}
        \comm{\hat{B}^{UR}_i}{\hat{H}} = -i\hbar 
    \end{equation}
    
    \item[] Ultra-Relativistic Boosts / Ultra-Relativistic Boosts\begin{equation}
        \comm{\hat{B}^{UR}_i}{\hat{B}^{UR}_j} = 2i\hbar(\frac{v^2}{c^2})a_{UR}[\hat{x}_i -\hat{x}_j]
    \end{equation} 
    
\end{enumerate}

\subsubsection{Non-Relativistic Cases: Schrödinger and Pauli}

We now proceed to derive the central non-relativistic operator, followed by the Schrödinger-type equation. The commutation with the non-relativistic boost operator imposes a structural constraint on the form of the equation,
\begin{equation}
    \comm{\hat{A}}{\hat{B}^{NR}_k} = 0 \xrightarrow{} \hat{A} \propto \delta_{ij}\hat{p}_i \hat{p}_j  + \hat{H}
\end{equation}
Due to the distinction between rotations and boosts, the operator exhibits a purely quadratic dependence on the spatial momentum operators $\hat{p}_i$. The Galilean boost symmetry restricts the temporal translation operator within $\hat{A}$, to a linear form.
We now express the non-relativistic central operator explicitly as: 
\begin{equation}
    \hat{A}^{NR}= a_{NR}(\alpha' \beta_i)(\alpha' \beta^i)\hat{p}_i \hat{p}^i + (\alpha' \beta_0)\hat{H} + \lambda I
\end{equation} 
Here, $\lambda$ is an additive constant that does not affect the internal structure of the operator. This grouping is consistent with the property $\{\alpha' \beta_i, \alpha' \beta_j \}= 0 \space $ for $(i \neq j)$, as discussed in Section 3.1.

The properties of the elements accompanying the symmetry generators arise from their commutation with the non-relativistic symmetry operator $\mathcal{\hat{S}}^{NR}$ (equation 25), and are as follows:

\begin{enumerate}
    \item[] Rotations: 
    \begin{equation}
    \begin{aligned}
    \comm{\alpha^i \beta^j}{(\alpha' \beta^k)^2} = 0  \\
    \comm{\alpha^i \beta^j}{\alpha' \beta^0} =0 \\
    \end{aligned}
    \end{equation}
    \item[] Boosts no relativistic:  
    \begin{equation}
    \begin{aligned}
    \comm{\alpha^0 \beta^i}{ (\alpha' \beta^k)^2} = 0 \\
    \comm{\alpha^0 \beta^i}{\alpha' \beta_0} = 0 \\
    \end{aligned}
    \end{equation}
    \item[] Translations: 
    \begin{equation}
    \begin{aligned}
    \comm{\alpha' \beta^\rho}{(\alpha' \beta^i)^2} = 0 \\
    \comm{\alpha' \beta^\rho}{\alpha' \beta^0} = 0 \\
    \end{aligned}
    \end{equation}
\end{enumerate}
 
Therefore, the terms $\alpha' \beta_0$ and $(\alpha'\beta^i)^2$, may be scalars or diagonal matrices (i.e., the identity matrix scaled by a constant). Now, applying Statement 2 and assuming $\alpha' \beta_0 = (\alpha'\beta^i)^2$, derives:
\begin{equation}
    \comm{\hat{A}}{\mathcal{\hat{D}}} \ket{\Phi_c}=0 \longrightarrow  [\alpha^i \beta^i  \hat{p}_i \hat{p}_i-\frac{\alpha_0 \beta_0}{2a_{NR}} \hat{H}] \ket{\Phi_c} = 0
\end{equation} 

Using the same simplification as in the relativistic case, namely $\alpha_\mu \beta_\mu = \mathbf{1}$, —a property that will also be reused in the ultra-relativistic case— we obtain the Schrödinger equation:\begin{equation}
    [\hat{p}_i \hat{p}^i - {2a_{NR}}\hat{H}] (\mathbf{1}\ket{\Phi_c}) = 0
\end{equation} 

In contrast, the Pauli equation without potential is a relativistic approximation that emerges when applying the symmetric transformation $\hat{D}_i = \frac{r_i}{c}\hat{B}^{NR}_i $, with $r_i = r_i(x)$ being a position-dependent function, to each $\hat{p}_i$ and $\hat{x}_i$ via an adjoint Galilean boost. This transformation ensures that the uncertainty relation remains invariant under the adjoint action (Gauge transformation).

\begin{equation}
\begin{aligned}
\hat{p}_i &\longrightarrow e^{\hat{D}_i} \hat{p}_i e^{-\hat{D}_i} = \hat{p}_i -\frac{q}{c}\phi_i \\
\hat{x}_i &\longrightarrow e^{\hat{D}_i} \hat{x}_i e^{-\hat{D}_i} = \hat{x}_i
\end{aligned}
\end{equation}
Given that $\comm{\phi_i}{ \hat{p}_i} = 0$, it follows that $\comm{\phi_i}{\hat{B}^{NR}_i} = 0$ in this case, and defining $r_i(x) = \frac{q}{i\hbar} \phi_i(x)$, where $\phi_i$ are the components of the vector potential $\vec{A} = (\phi_1,\phi_2,\phi_3)$ and $q$ is the electric charge of the particle. Assuming once again $\alpha_\mu \beta_\mu = \mathbf{1}$, and applying Statement 2, we obtain the Pauli equation. 

\begin{equation}
     [(\alpha'\beta^i) (\alpha'\beta_i) (\hat{p}_i -\frac{q}{c}\phi_i) (\hat{p}^i -\frac{q}{c}\phi^i) - \frac{\alpha'\beta_0 }{2a}\hat{H}] (\mathbf{1}\ket{\Phi_c}) = 0
\end{equation}
Note that this maps perfectly onto the Pauli equation when we take $\alpha'\beta_i = \sigma_i$ $\alpha'\beta_0 = I_2$, and express the operators in their differential form. The algebraic derivation of the Pauli equation from the non-relativistic algebra via a Galilean boost transformation offers a straightforward explanation of why its structure is not relativistic—it is merely an approximation valid for particles with half-integer spin. See, for instance, an algebraic construction in \cite{Wilkes2019}, where the Pauli current is derived from spin identities within the framework of Lévy-Leblond theory. Furthermore, in the limit $c \xrightarrow[]{} \infty$, we recover the non-relativistic Schrödinger structure.

\subsubsection{Ultra-Relativistic Case: Carroll–Schrödinger} 

We now explore a particular case in which the system exhibits translational symmetries but lacks rotational symmetries. Our main goal is to construct an equation analogous to the scalar Schrödinger equation, but with a linear dependence on the spatial translation operators and a second-order dependence on the temporal translation operator. This construction allows us to study systems that, while not strictly relativistic, display intriguing properties that bring them closer to that regime. In this context, the system is anisotropic (in the scalar case), and the central Casimir operator $\hat{A}_c$ does not commute with spatial rotations, assuming from equation (26) that: $(\alpha^i \beta^j\hat{M}_{i j} + \alpha' \beta'/c^2) \xrightarrow{} 0$. 
Then, \begin{equation}
    \comm{\hat{A}_c}{\hat{B}^{UR}_k} = 0 \xrightarrow{} \hat{A}_c \propto \sum_i \hat{p}_i   + \hat{H}^2
\end{equation}

we can now write:
\begin{equation}
    \hat{A}_c= (\frac{c^2}{v^2a^*_{UR}}) \alpha' \beta^i \hat{p}_i + (\alpha' \beta^0)^2\hat{H}^2+ \lambda_c \hat{I}
\end{equation}

Then, applying Statement 2:
\begin{equation}
    [(\frac{c^2}{2v^2a^*_{UR}}) \alpha' \beta^i \hat{p}_i - (\alpha' \beta^0)^2\hat{H}^2]\ket{\Phi_c}=0 
\end{equation}
with $\alpha' \beta^\mu$ treated as scalars. This equation describes the dynamics of the system in the scalar ultra-relativistic limit, where the contributions from the translation operators are coupled in a non-trivial way.
The differential form is obtained by substituting the operators, $\hat{p}_i = -i\hbar \partial_i$. y $\hat{H} = i\hbar \partial_t$, thus representing the Carroll–Schrödinger equation as: 

\begin{equation}
   [\frac{ -i\hbar}{2} \hat{n} \cdot \grad + a^*_{UR}\hbar^2\partial_t^2] \ket{\Phi_c}=0
\end{equation}
We define the vector $\hat{n} = (\alpha' \beta^1,\alpha' \beta^2, \alpha' \beta^3) \in \mathbb{R}^3 $ and set $\alpha' \beta^0 = c/v$. To illustrate the type of solutions admitted by this differential equation, we consider the linear case $\ket{\Phi_c} = \ket{\Phi_c}(x,t)$. Its analytical solution resembles the inverted kernel of the heat equation, with domain $(x,t) \in \{ \mathbb{R} \setminus \{0\}\ \cross \mathbb{R} \}$. By choosing $a^*_{UR} = -\frac{1}{4(mc^3)}$ the solution becomes: 
\begin{equation}
    \ket{\Phi_c} =  c\sqrt{\frac{mc}{2\pi i\hbar x}} exp({\frac{imc^3 t^2}{2\hbar x}})
\end{equation}
This satisfies the limiting condition, $ \lim_{x \xrightarrow{}0}\ket{\Phi_c}(x,t)=\delta(t)$.
The solution behaves as a Gaussian distribution proportional to the transition amplitude in Carroll–Schrödinger field theory, indicating spatial localization. This result is consistent with the one dimensional Schrödinger equation under the exchange $x \longleftrightarrow ct$ and highlights the interchangeable duality between spatial and temporal dispersion in ultra-relativistic and no-relativistic regimes in the scalar case.

\subsubsection{Restoration of Rotational Symmetries: Carroll–Schrödinger–Pauli} 
 
To construct a general operator that restores rotational symmetries, it is necessary to assume that the first term of $\mathcal{\hat{S}}^{UR}$ does not vanish, and therefore the operator $\hat{A}_{c}$ is not scalar. We use the anticommutation relation $\{\alpha' \beta_i, \alpha'\beta_j \} = 0$ for $(i \neq j)$, as in the non-relativistic case. 
It is known that in the scalar case, the rotation generator $\hat{M}_{i j }$ does not commute with $\hat{A}_c$. By introducing the coupling $\alpha' \beta' /c^2 = \sum E_{i j}$, we mimic the Dirac case, recovering the commutation structure. The condition becomes:
\begin{equation}
     (\alpha' \beta^i) \comm{\hat{p_i}}{\hat{M}_{k j}} + \comm{(\alpha' \beta^i)}{E_{k j}}\hat{p}_i = 0 
\end{equation}

The properties of the elements accompanying the symmetry generators in equation (36) arise from their commutation with $\mathcal{\hat{S}}^{UR}$ (equation 26). These elements satisfy:

\begin{enumerate} 
    \item[] Rotations: \begin{equation}
    \begin{aligned}
    \comm{\alpha' \beta^k}{\alpha^i \beta^j} = 0  \\
    \comm{(\alpha' \beta^0)^2}{\alpha^i \beta^j} = 0 \\
    \end{aligned}
    \end{equation} 
       
        This ensures the commutation: $\comm{\hat{A}}{\alpha_i \beta_j(\hat{M}_{i j} + E_{i j})} = 0$
    \item[] Ultra-relativistic boosts: \begin{equation}
        \begin{aligned}
        \comm{\alpha' \beta^i}{\alpha^0 \beta'} = 0  \\
        \comm{\alpha' \beta^i}{\alpha^i \beta'} = 0 \\ 
        \comm{(\alpha' \beta^0)^2}{\alpha^0 \beta'} = 0 \\ 
        \comm{(\alpha' \beta^0)^2}{\alpha^i \beta'} = 0
        \end{aligned}
        \end{equation}  
    \item[] Translations: \begin{equation}
            \comm{(\alpha' \beta^0)^2}{\alpha' \beta^\rho} = 0
    \end{equation}

\end{enumerate} 
Due to the antisymmetry induced by the coupling with the rotation operators, the term $E_{i j}$ can be written as:
\begin{equation}
    E_{i j} = -\frac{i\hbar}{2}\comm{\alpha'\beta_i}{\alpha'\beta_j}
\end{equation}
This restores the full set of 10 symmetries, as in previous cases within Minkowski spacetime. Since rotational symmetry is recovered, a spin interpretation becomes viable in the ultra-relativistic regime. The most suitable candidates for the elements $\alpha'\beta^i$ are the Pauli matrices, as in the non-relativistic case.
Finally, by setting $\alpha' \beta^0 = I_2(c/v) $, equation (45), derived after Statement 2, becomes:  
\begin{equation}
    [\sigma^i \hat{p}_i - 2a^*_{UR} \hat{H}^2] (I_2\ket{\Phi_c}) = 0
\end{equation}
To better understand this equation, we consider the solution  \(\ket{\Phi_c} = u_+ e^{i (\mathbf{k} \cdot \mathbf{r} - \omega t)} + u_-e^{-i (\mathbf{k} \cdot \mathbf{r} - \omega t)}\), where \(u_+\) and \(u_-\) are constant spinors, provided that:
\begin{itemize}
    \item \(u_+\) is a spinor with spin antiparallel to \(\mathbf{k}\), i.e \(\vec{\sigma} \cdot \mathbf{k} u_+ = -|\mathbf{k}| u_+\).
    \item \(u_-\) is a spinor with spin parallel to \(\mathbf{k}\), i.e, \(\vec{\sigma} \cdot \mathbf{k} u_- = +|\mathbf{k}| u_-\).
    \item Both terms satisfy the dispersion relation:
    \begin{equation}
    \omega = \sqrt{\frac{2 m c^3 |\mathbf{k}|}{\hbar}}.
    \end{equation}
\end{itemize}

Physically, this solution describes a superposition of two plane waves with opposite momenta (\(\hbar \mathbf{k}\) y \(-\hbar \mathbf{k}\)), and spins correlated with their propagation directions. The first term in equation (55) is related to the helicity operator, which measures the projection of a particle’s spin along its momentum direction—a concept widely used in particle collisions involving arbitrary spin \cite{JacobWick1959}). Such solutions may represent states in systems with spin–momentum coupling, as found in topological materials \cite{HasanKane2010}, or in effective field theories with non-standard temporal dynamics governed by the dispersion relation $\omega \propto \sqrt{|\mathbf{k}|}$, in contrast to relativistic systems $\omega \propto {|\mathbf{k}|}$ or non-relativistic particles $\omega \propto k^2$. 

To complete the analogy with the Pauli equation, we now apply the same adjoint transformation used at the end of Section 3.2.1, but this time with the ultra-boost, $\hat{D}_i = w_i\hat{B}^{UR}_i$. This transformation preserves the Heisenberg uncertainty relation, ensuring that the quantum structure remains invariant under the ultra-relativistic symmetry as well. Thus,
\begin{equation}
\begin{aligned}
\hat{p}_i &\longrightarrow e^{\hat{D}_i} \hat{p}_i e^{-\hat{D}_i} = \hat{p}_i - (Qa^*_{UR}) \phi_i \hat{H} + (\frac{c^2}{v^2}) \frac{Q^2}{2} a^*_{UR} \phi^2_i\\
\hat{x}_i &\longrightarrow e^{\hat{D}_i} \hat{x}_i e^{-\hat{D}_i} = \hat{x}_i
\end{aligned}
\end{equation}
with the function  $w_i(x) = \frac{c^2}{v^2} \frac{Q}{2i \hbar} \phi_i(x) $, where $Q$ is a real constant, potentially interpreted as the electric charge of the system. The following commutation relations are satisfied, $\comm{\hat{p}_i}{\phi_i} = 0$ and $\comm{\hat{B}^{UR}_i}{\phi_i} = 0$. The only additional condition required to derive equation (56) is: $\comm{\hat{H}}{\phi_i}=0$, which is as well, implicitly satisfied in the Pauli equation (43) as long as the vector potential is not an explicit function of time. 
From equation (56), we now write the Carroll–Schrödinger–Pauli equation, which generalizes the Pauli structure to ultra-relativistic regimes with spin–momentum coupling and a non-standard dispersion relation. 
\begin{equation}
     [\sigma^i (\hat{p}_i -Qa^*_{UR} \phi_i \hat{H} + (\frac{c^2}{v^2}) \frac{Q^2}{2} a^*_{UR} \phi^2_i)- 2a^*_{UR} \hat{H}^2] (I_2\ket{\Phi_c}) = 0
\end{equation}
Note that in the limit $a^*_{UR} \xrightarrow[]{} 0$, we obtain:
\begin{equation}
    \sigma^i \hat{p}_i \ket{\Phi_c} = 0
\end{equation}
where $\ket{\Phi_c}$ is one of the possible solutions to the Weyl equation for massless particles with spin $s=1/2$. Nevertheless, the parameter $a^*_{UR}$ has units of $(kg *m^3/s^3)^{-1}$, which implies that equation (58) can be approximated in regimes where the system is highly energetic or possesses an extremely large mass.

\section{Discussion}

The derivation of a Carroll–Schrödinger-type equation in the regime where $c \xrightarrow{} 0$ raises several theoretical aspects worth discussing. First, the fact that rotational symmetry is broken in the scalar case, while invariance under ultra-boosts is preserved, suggests an anomaly in which the symmetry generators do not form a Poincaré-type subalgebra involving translations, rotations, and boosts.

This could have significant implications for the study of effective theories with inherent anisotropy \cite{Soluyanov2015}. Nevertheless, the validity of the Carrollian limit still requires interpretation regarding the consistency of temporal evolution, which in this model is governed by quadratic terms. The extension with spin, obtained through the introduction of Pauli matrices, suggests a formal analogy with the Dirac construction, although the physical framework is distinct. Here, spin does not emerge as a consequence of a relativistic symmetry representation, but rather as a structure added for internal consistency of the equation to recover rotations—similar to the non-relativistic Pauli approximation. This raises the question of whether a more general symmetry principle exists—possibly related to a non-linear or deformed extension of the Poincaré–Carroll algebra—that could naturally derive these structures from a modified metric  $\eta_{\mu v} \xrightarrow{} g_{\mu v}$, a generalized uncertainty principle (GUP). In particular, it would be interesting to investigate whether the operator space $Z_{\alpha \beta}$ admits a geometric realization or can be interpreted as a representation of a non-associative super-algebra relevant to models in quantum gravity theories. See: \cite{Casadio2014} \cite{Chemisana2023} \cite{Cremaschini2021} \cite{Das2008} \cite{Wagner2021}. Likewise, an extended algebra with corrections derived from perturbations in the Heisenberg–Weyl uncertainty due to noisy measurement experiments or open quantum systems, etc. See: \cite{Ozawa2002} \cite{Ozawa2003} \cite{Busch2007} \cite{Rozema2012}  \cite{Woods2013} \cite{Pienaar2013} \cite{Busch2013}. In order to derive a family of free-particle equations of motion for each modified HUP from our algebraic grouping. However, one of the limitations is the assumption of compatibility with the type I Von Neumann algebra, which applies only to discrete systems and fails in local algebraic theories of observables in quantum field theory and in quantum statistical systems. Therefore, to approach quantum gravity theories with infinite degrees of freedom, it is necessary, for example, to move to the type III Von Neumann algebra.

Furthermore, the modification in the dispersion relation that emerges in the ultra-relativistic model may be interpreted as a sign of non-standard wave behavior. It would be desirable to explore whether this behavior can manifest in concrete physical phenomena, such as the propagation of particles in anisotropic media for bosonic cases, and in isotropic fermionic systems.

\section{Conclusion}

We have presented an algebraic formulation for deriving equations of motion in quantum mechanics based on symmetry principles and the minimal postulate of uncertainty. Through the construction of conjugate operators in phase space and the imposition of commutation with a central Casimir-type operator, we have shown that it is possible to recover known quantum dynamics—both relativistic and non-relativistic—and to generate new structures. In particular, the ultra-relativistic Carroll–Schrödinger–Pauli case stands out as a notable example of this construction, proposing a matrix-type free particle equation with spin. This enables an extension and approximation within a distinct quantum dynamic, demonstrating the versatility of the proposed framework. 

\section{Acknowledgments}
This article was inspired by the work of the PhD. Nikolay Sukhomlin.

\bibliographystyle{plain}

\typeout{get arXiv to do 4 passes: Label(s) may have changed. Rerun}

\end{document}